\documentclass[aps,floatfix,prd,nofootinbib,notitlepage,10pt]{revtex4-1}

\usepackage{graphicx}
\usepackage{epic}
\usepackage{eepic}
\usepackage{latexsym}
\usepackage{amssymb,amsmath}

\newcommand{\eq}[1]{(\ref{#1})}
\newcommand{\be}{\begin{equation}}
\newcommand{\ee}{\end{equation}}
\newcommand{\bea}{\begin{eqnarray}}
\newcommand{\eea}{\end{eqnarray}}

\newcommand{\hs}[1]{\hspace{#1 mm}}

\newcommand{\df}{\dot{\phi}}
\newcommand{\zo}{\zeta_k^{(0)}}
\newcommand{\cco}{\cc_k^{(0)}}

\newcommand{\lf}{\left<}
\newcommand{\rg}{\right>}

\newcommand{\CF}{{\cal F}}

\newcommand{\CH}{{\cal H}}

\newcommand{\mm}{\tilde{m}}

\def\a{\alpha}
\def\b{\beta}
\def\cc{\gamma}
\def\C{\Gamma}
\def\d{\delta}

\def\e{\epsilon}

\def\f{\phi}
\def\fr{\frac}
\def\F{\Phi}
\def\vf{\varphi}

\def\l{\lambda}

\def\s{\sigma}
\def\S{\Sigma}

\def\th{\theta}

\def\z{\zeta}

\def\o{\omega}

\def\del{\partial}

\let\bm=\bibitem
\def\nn{\nonumber}

\begin{document}

%\large

\title{Entropy mode loops and cosmological correlations during perturbative reheating} 

\author{Ali Kaya}

\email[]{ali.kaya@boun.edu.tr}

\author{Emine Seyma Kutluk}

\email[]{seymakutluk@gmail.com} 

\affiliation{Bo\~{g}azi\c{c}i University, Department of Physics, 34342, Bebek, \.{I}stanbul, Turkey}

\date{\today}

\begin{abstract}

Recently, it has been shown that during preheating the entropy modes circulating in the loops, which correspond to the inflaton decay products, meaningfully modify the cosmological correlation functions at superhorizon scales. In this paper, we determine the significance of the same effect when reheating occurs in the perturbative regime. In a typical two scalar field model, the magnitude of the loop corrections are shown to depend on several parameters like the background inflaton amplitude in the beginning of reheating, the inflaton decay rate and the inflaton mass. Although the loop contributions turn out to be small as compared to the preheating case, they still come out  larger than the loop effects during inflation. 

\end{abstract}

\maketitle

\section{Introduction}

According to inflation, the basic cosmological observables are mainly fixed by the tree level quantum field theory amplitudes. Obviously, to have a complete understanding of the inflationary predictions one must determine the loop corrections \cite{w1} or explore possible nonperturbative effects \cite{n1,n2}. It turns out that loop effects during inflation become small \cite{w2}. Besides, in single field models the curvature  perturbation $\z$ is shown to be conserved at superhorizon scales to all loop orders \cite{loop1,loop2} (see also \cite{ekw}). Although the loop amplitudes are plagued by IR divergences and a careful treatment is needed to extract the physical observables (see e.g. \cite{ir0,ir1,ir2,ir22,ir3}), the loop corrections during inflation are most likely negligible, at least in single scalar field models (see also \cite{e1,e2,e3,e4,e5,e6} for some other  work on loop effects in de Sitter space).   

On the other hand, it has been shown in \cite{a1,a2} that the loop effects during preheating can significantly modify the cosmological correlation functions at superhorizon scales. This somehow surprising result holds due to (i) the presence of the entropy perturbations (ii) nonlinearities and (iii) exponential growth of the preheating mode functions. It is known that the superhorizon conservation of the curvature perturbation $\zeta$ breaks down in the presence of entropy perturbations (see e.g. \cite{mfb}). One may think that  the entropy modes produced by the inflaton decay during preheating cannot affect the superhorizon modes since they have in general short wavelengths characterized by the instability bands. However,  the nonlinearities introduced by the interactions give rise to mode-mode coupling and consequently  the  Fourier modes do not evolve independently. This allows short scale entropy modes to affect the long wavelength adiabatic modes.  Furthermore, the mode functions of the preheating scalar corresponding to the inflaton decay are subject to exponential growth \cite{reh1,reh2,reh3,reh4}. Thus,  quantum corrections having these modes circulating in the loops are greatly enhanced. 

The possibility that the superhorizon metric perturbations are modified during preheating has been pointed out in \cite{mph1,mph2,mph3}, but that scenario turned out to be working only for the massless preheating scalar models \cite{mph4,mph5,mph6} because of the suppression of the massive mode functions during inflation \cite{mpk1,mpk2,mpk3}. This earlier discussion has been mainly  carried out at the linearized level 
and it is shown in \cite{a1,a2} that the suppression of the massive mode functions during inflation are compensated in the loop integrals during preheating. As we will see below, the same happens for the perturbative reheating, which avoids the suppression problem pointed out in \cite{mpk1,mpk2,mpk3}. 

In this paper, we would like to determine the contributions of the entropy mode loops  to the scalar and the tensor power spectra during perturbative reheating, where the decay of the inflaton can be described as the slow particle creation by the oscillating inflaton background. These corrections are expected to be smaller as compared to the preheating case since the corresponding energy scale is much lower and the mode functions do not grow exponentially. Nevertheless, the decay process takes a lot longer, which would strengthen the effect according to the in-in perturbation theory as we will point out in section \ref{s4}. In any case, we find that the corrections are generically very small compared to preheating but they are still significantly larger than the loop corrections during inflation. 

The plan of the paper is as follows: In the next section we introduce the model containing two scalar fields, the inflaton $\f$ and the reheating scalar $\chi$, and we specify the background evolution.  Then, following \cite{ent}, the fluctuations of the $\chi$-field are identified as the entropy perturbations. In section \ref{s3}, we determine the linearized mode functions in the gauge $\z=0$ and study the $\chi$-particle creation effects in detail from the  Bogoliubov coefficients. In section \ref{s4}, we  determine the cubic interactions involving two $\chi$-fields and use the in-in perturbation theory to calculate the one-loop corrections to the scalar and the tensor power spectra when the $\chi$-modes are circulating in the loops. We also elaborate on the regularization and the renormalization of the loop contributions. In section \ref{s5}, we conclude by summarizing our results and pointing out  future directions.

\section{The Model and the background} \label{s2}

We consider a two scalar field model containing the inflaton $\f$ and the reheating scalar $\chi$. The $\chi$-field does not play a role during inflation and it is mainly responsible for the inflaton decay and reheating. The scalars are minimally coupled to gravity and as usual the dynamics of the system is governed by the Einstein-Hilbert action, which can be written in the ADM from as 
\be\label{a}
S=\fr12 \int \sqrt{h} \left[NA+\fr{B}{N}\right],
\ee
where $N$ and $N^i$ are the standard lapse and shift functions  of the metric
\be
ds^2=-N^2dt^2+h_{ij}(dx^i+N^i dt)(dx^j + N^j dt),
\ee
$K_{ij}=\fr12(\dot{h}_{ij}-D_i N_j-D_j N_i)$, $K=h^{ij}K_{ij}$, $D_i$ is the derivative operator of $h_{ij}$ and 
\bea
&&A=M_p^2R^{(3)}-2V-h^{ij}\del_i\phi\del_j\f-h^{ij}\del_i\chi\del_j\chi,\label{aa}\\
&&B=M_p^2K_{ij}K^{ij}-M_p^2K^2+(\dot{\f}-N^i\del_i\f)^2+(\dot{\chi}-N^i\del_i\chi)^2.\label{bb}
\eea
Here, the reduced planck mass is defined as $M_p^{-2}=8\pi G$. After inflation, $\f$ is assumed to be oscillating about  its minimum, thus the potential during reheating can be taken as
\be
V=\fr12 m^2\f^2+\fr12 \mm^2\chi^2+\fr12 \s \f\chi^2 ,\label{p1}
\ee
where $m$ and $\mm$ are the corresponding masses and $\s$ characterizes the inflaton decay rate by the cubic interaction. The  cubic interaction term, which is responsible for the inflaton decay, will be treated perturbatively and therefore it is enough to have a local stable vacuum around $\f=\chi=0$. 

The background of $\chi$ vanishes until the created $\chi$-particles start affecting the background evolution. As it is well known, the inflaton oscillations can be characterized by an average equation of state $P=0$ (one has $\rho=-P$ and $\rho=P$ for $\dot{\f}=0$ and $\f=0$, respectively). We assume $m\gg H$ and in that case the background fields can be determined  as 
\bea
&&h_{ij}=a(t)^2\d_{ij},\nn\\
&&\f(t)=\F\sin(mt),\label{back}\\
&&N=1,\hs{3} N^i=0, \hs{3}\chi=0,\nn
\eea
where
\be\label{rd}
\F\simeq \F_i\left(\fr{t_i}{t}\right),\hs{5}a(t)\simeq \left(\fr{t}{t_i}\right)^{2/3}.
\ee
We define $t_i$ and $t_f$ to denote the beginning of reheating and the end of the stage after which the backreaction effects become important (in the next section we fix $t_f$ in terms of other parameters). Therefore, the background solution \eq{back} is valid in the interval $(t_i,t_f)$.  The background Friedmann equation  $H^2=\fr{1}{6M_p^2}(\df^2+m^2\f^2)$, which approximately takes the following from 
\be
H^2\simeq \fr{4}{9t^2} \simeq\fr16 \fr{m^2\F^2}{M_p^2}, \label{t}
\ee
can be used to relate the cosmic time $t$ to the inflaton amplitude $\F$. We fine tune $t_i$ so that 
\be\label{smti}
\sin(mt_i)=1
\ee
and thus $\F_i$ denotes the initial value of the inflaton in the beginning of reheating, which can be smoothly matched to the inflationary stage. Note that \eq{smti} is consistent with \eq{t} since $mt\gg1$. The value of $\F_i$,  which we take as an input parameter, depends on the inflationary stage and our assumption $m\gg H$ implies $\F_i\ll M_p$. 

As we will discuss in the next section, the perturbative reheating process can be viewed  as $\f$-particles of mass $m$ decaying into doublets of $\chi$-particles with energies $m/2$. We will assume that $m\gg\mm$ so that the mass of the $\chi$-field is negligible in the whole decay process. It turns out that (see the next section) for the perturbative regime to be valid,  the parameters must obey\footnote{If $\chi$ is required to be massive enough during inflation to suppress its own fluctuations, one needs to impose $\mm>H$ or $\s\F_i>H^2$. For $\mm=0$, this would give a lower limit for $\s$ as $\s M_p>m^2\F_i/M_p$. Together with \eq{c1}, this implies $m^2>\s M_p>m^2\F_i/M_p$. \label{foot1}}
\be
\s \sqrt{\F_i M_p}\ll m^2,\hs{5} \s M_p\ll m^2.\label{c1}
\ee
Since $\F_i\ll M_p$, the second condition is more restrictive and it implies the first one. When \eq{c1} is satisfied, it is possible to calculate the Bogoliubov coefficients corresponding to the $\chi$-particle creation on the background \eq{back}. Using these Bogoliubov coefficients, one may find  that the backreaction starts when the inflaton amplitude reduces to (see the next section)
\be
\F_f\simeq 10^{-2}\fr{\s^2M_p}{m^2}.\label{pf}
\ee
The corresponding time $t_f$ can be determined from \eq{t}. Assuming that the $\chi$-particles are thermalized instantaneously at $t_f$, the reheating temperature can be found as
\be
T_r\simeq 0.1\sqrt{\fr{M_p\s^2}{m}}.\label{rt}
\ee
This is exactly the reheating temperature in the perturbative regime with the decay rate $\C=\s^2/m$ that corresponds to the cubic interaction in \eq{p1}. 

As discussed in \cite{ent}, in a two field model like the one considered in this paper, the adiabatic field $\S$ and the entropy perturbation $\d s$ can be defined as 
\bea
&&\dot{\S}=(\cos\th)\df+(\sin\th)\dot{\chi},\\
&&\d s=(\cos\th)\d\chi-(\sin\th)\d \f,\nn
\eea
where
\be
\cos\th=\fr{\df}{\sqrt{\df^2+\dot{\chi}^2}},\hs{3} \sin\th=\fr{\dot{\chi}}{\sqrt{\df^2+\dot{\chi}^2}}. 
\ee
From \eq{back}, one sees that $\S=\f$ and $\d s =\d\chi$, thus $\d\f$ and $\d\chi$ become the adiabatic and the entropy perturbations, respectively. 

\section{The linearized modes} \label{s3}

Let us now consider the evolution of the cosmological perturbations in this model. As usual, the scalar and the tensor fluctuations can be parametrized  as
\bea
&&h_{ij}=a^2e^{2\zeta}(e^\cc)_{ij},\nn\\
&&\phi=\phi(t)+\vf,\\
&&\chi=0+\chi.\nn
\eea
Note that $\chi$ now denotes the fluctuation field since the corresponding background value vanishes. The gauge freedom of the infinitesimal coordinate transformations are completely fixed by imposing
\be\label{gauge}
\z=0,\hs{5}\del_i\cc_{ij}=0,\hs{5} \cc_{ii}=0.
\ee
For convenience, we prefer to set $\z=0$ since $\z$ becomes an ill defined variable during reheating, i.e. it blows up at times $\df=0$ and these spikes must be smoothed out as discussed in \cite{a2}. To first order in fluctuations, the lapse and the shift can be solved as \cite{mal}  
\be\label{ls}
N=1+\fr{\dot{\f}}{2HM_p^2}\vf,\hs{5}N^i=\d^{ij}\del_j\psi,
\ee
where 
\be
\del_i\del^i\psi=-\fr{1}{4HM_p^2}\left[\fr{m^2\f^2\df}{HM_p^2}\vf+m^2\f\vf+2\df\dot{\vf}
\right].
\ee
Since we will only deal with the cubic interaction terms, it is enough to determine $N$ and $N^i$ to first order \cite{mal}. Using these solutions, it is a straightforward exercise to obtain the following quadratic actions:
\bea
&&S_\vf^{(2)}=
\fr12 \int a^3\left[\dot{\vf}^2-\fr{1}{a^2}(\del\vf)^2-m^2\left(1+\fr{\f\df}{2HM_p^2}\right)^2\vf^2-\fr{\df^2}{HM_p^2}\vf\dot{\vf}\right],\nn\\
&&S_\chi^{(2)}=\fr12\int a^3\left[\dot{\chi}^2-\fr{1}{a^2}(\del\chi)^2-\mm^2\chi^2-\s\f \chi^2\right],\label{qac}\\
&&S_\cc^{(2)}=\fr18\int a^3\left[\dot{\cc}_{ij}^2-\fr{1}{a^2}(\del\cc_{ij})^2\right].\nn
\eea
For quantization, one may introduce the ladder operators as 
\bea
&&\vf=\fr{1}{(2\pi)^{3/2}}\int d^3k\, e^{i\vec{k}.\vec{x}}\,\vf_k(t) a_{\vec{k}}+h.c.\label{m1}\\
&&\chi=\fr{1}{(2\pi)^{3/2}}\int d^3k\, e^{i\vec{k}.\vec{x}}\,\chi_k(t) \tilde{a}_{\vec{k}}+h.c.\label{m2}\\
&&\cc_{ij}=\fr{1}{(2\pi)^{3/2}}\int d^3k\, e^{i\vec{k}.\vec{x}}\,\cc_k(t) \e_{ij}^s \tilde{a}^s_{\vec{k}}+h.c.\nn
\eea
where $s=1,2$, the polarization tensor $\e^s_{ij}$ obeys
\be
k^i\e^s_{ij}=0,\hs{5} e^s_{ii}=0,\hs{5} \e^s_{ij}e^{s'}_{ij}=2\d^{ss'}, 
\ee
and  the creation-annihilation operators satisfy the usual relations, e.g. $[a_k,a^\dagger_{k'}]=\d^3(k-k')$. The linearized mode equations are given by  
\bea
&&\ddot{\vf}_k+3H\dot{\vf}_k+\left[m^2+\fr{2m^2\f\df}{HM_p^2}+\fr{3\df^2}{M_p^2}-\fr{\df^4}{2H^2M_p^4}+\fr{k^2}{a^2}\right]\vf_k=0,\nn\\
&&\ddot{\chi}_k+3H\dot{\chi}_k+\left[\s\f+\fr{k^2}{a^2}+\mm^2\right]\chi_k=0,\label{le}\\
&&\ddot{\cc}_k+3H\dot{\cc}_k+\fr{k^2}{a^2}\cc_k=0.\nn
\eea
For the canonical commutation relations to hold, the mode functions must obey the Wronskian conditions 
\bea
&&\vf_k\dot{\vf}_k^*-\vf_k^*\dot{\vf}_k=\fr{i}{a^3},\nn\\
&&\chi_k\dot{\chi}_k^*-\chi_k^*\dot{\chi}_k=\fr{i}{a^3},\label{w}\\
&&\cc_k\dot{\cc}_k^*-\cc_k^*\dot{\cc}_k=\fr{4i}{a^3M_p^2}.\nn
\eea
Naturally, deep inside the horizon during inflation the Bunch-Davies vacua are chosen for each field and this choice fixes the mode functions uniquely. 

For the modes $\vf_k$ and $\cc_k$, we only need the superhorizon evolution. Neglecting $k^2/a^2$ terms in \eq{le}, it is an easy exercise to obtain the two linearly independent solutions for $\vf_k$ and $\cc_k$ as 
\be\label{sh}
\vf_k\simeq \fr{\df}{H}\left[\zo+c_k f(t)\right],  \hs{5} \cc_k\simeq \left[\cco+d_k g(t)\right],
\ee
where $\zo$, $\cco$, $c_k$ and $d_k$ are constants and 
\be\label{fg}
\fr{df}{dt}=\fr{H^2}{a^3\df^2}, \hs{5}  \fr{dg}{dt}=\fr{1}{a^3}.
\ee
Using \eq{back}, $f$ and $g$ can be solved approximately as
\be
f\simeq\fr{H^2\sin(mt)}{a^3\Phi m^2\df},\hs{5}g\simeq\fr{2}{3Ha^3}.\label{fgs}
\ee
Although $f$ diverges when $\df=0$, the product $\df f$, which actually appears in \eq{sh}, is well defined. Substituting \eq{sh} in \eq{w}, the Wronskian conditions can be seen to imply  
\be\label{n}
\zo c_k^* -\zo{}^* c_k=i,\hs{5} \cco d_k^* -\cco{}^* d_k=\fr{4i}{M_p^2}.
\ee
In \eq{sh}, while  $\zo$ and  $\cco$  give the ``constant'' modes,  the others correspond to the decaying pieces as the corresponding functions have $1/a^3$ factors in \eq{fgs}. 

The scalar and the tensor power spectra in the momentum space, i.e. $P_k^\vf$ and $P_k^\cc$, can be defined from the two point functions in the position space as follows 
\bea\label{sp}
&&\lf\vf(t,\vec{x})\vf(t,\vec{y})\rg=\fr{1}{(2\pi)^{3}}\int d^3k \,e^{i\vec{k}.(\vec{x}-\vec{y})}\,P_k^\vf(t),\\
\label{tp}
&&\lf\cc_{ij}(t,\vec{x})\cc_{kl}(t,\vec{y})\rg=\fr{1}{(2\pi)^{3}}\int d^3k \,e^{i\vec{k}.(\vec{x}-\vec{y})}\,P_k^\cc(t)\Pi_{ijkl},
\eea
where the polarization tensor $\Pi_{ijkl}$ is given by
\be\label{pt}
\Pi_{ijkl}=e^s_{ij}e^s_{kl}=P_{ik}P_{jl}+P_{il}P_{jk}-P_{ij}P_{kl},\hs{7}P_{ij}=\d_{ij}-\fr{k_ik_j}{k^2}.
\ee  
Note that  one has $\Pi_{ijkl}\Pi_{klmn}=2\Pi_{ijmn}$. Using \eq{sh} in \eq{sp}, the standard tree level results can be obtained as 
\be\label{3l}
P_k^{\vf(0)}(t)=\fr{\df(t)^2}{H(t)^2}|\zo|^2,\hs{5} P_k^{\cc(0)}(t)=|\cco|^2.
\ee
While $P_k^{\vf(0)}(t)$ is nearly constant during inflation, it starts oscillating during reheating. Note also that $\zo$ corresponds to the constant superhorizon curvature perturbation. 

Our aim is to calculate the 1-loop corrections to \eq{3l} by the $\chi$-modes circulating in the loops during reheating. For this, the evolution of $\chi_k$ must be determined in detail.  It is important to recognize that due to the cubic coupling in \eq{p1}, $\chi$ becomes effectively a massive field even for $\mm=0$.  Moreover, the momenta of the modes corresponding to the inflaton decay are of the order of $k_{phys}\simeq m/2\gg H$. Thus, the modes of interest evolve adiabatically during inflation and up to a constant phase the initial value of the corresponding mode functions in the beginning of reheating can be found as
\be\label{sup}
\chi_k(t_i)\simeq \fr{1}{\sqrt{2a(t_i)^3\o_k(t_i)}},\hs{5}\dot{\chi}_k(t_i)\simeq i\sqrt{\fr{\o_k(t_i)}{2a(t_i)^3}}.
\ee
As noted in \cite{mpk1,mpk2,mpk3}, the scale factors in \eq{sup} suppress the modes by the factor $e^{-3N/2}$ during inflation, where $N$ is the number of e-folds. As we will see in the next section, the scale factors, which are arising from the Hamiltonians and from the loop momentum integrals, have just enough powers to cancel the suppression in a loop correction. 

To determine the evolution of $\chi_k$ during reheating, we write it in the WKB form as 
\be\label{wkb}
\chi_k=\fr{1}{\sqrt{2a^3\o_k}}\left[\a_k e^{-i\th_k}+\b_k e^{+i\th_k}\right],
\ee
where 
\bea
&&\o_k^2=\mm^2+\fr{k^2}{a^2}+\s\f-\fr94 H^2-\fr32 \dot{H},\label{o}\\
&&\th_k(t)=\int_{t_i}^t \o_k(t')dt'.\nn
\eea
Note that since $H\simeq 2/(3t)$, the last two terms in \eq{o} actually cancel each other. From \eq{le}, the Bogoliubov coefficients can be seen to obey
\bea
&&\dot{\a}_k=\fr{\dot{\o}_k}{2\o_k}e^{2i\th_k}\b_k,\label{ab}\\
&&\dot{\b}_k=\fr{\dot{\o}_k}{2\o_k}e^{-2i\th_k}\a_k.\nn
\eea 
For the modes satsifying $\o_k\gg H$, which is the case of interest for us, the initial conditions can be fixed from \eq{sup} as
\be
\a_k(t_i)=1,\hs{5}\b_k(t_i)=0.\label{ic}
\ee
The evolution of the modes is now uniquely determined  by \eq{ab}. The Wronskian condition \eq{w} for $\chi_k$ implies $|\a_k|^2-|\b_k|^2=1$, which is preserved by the equations of motion \eq{ab}. 

In this model,  reheating occurs by the particle creation due to the  oscillating inflaton background $\f$ that appears in \eq{o}. This process is slow and thus can be treated perturbatively if  $|\a_k|\simeq 1$ and $|\b_k|\ll1$. In that case, $\b_k$ can be calculated by applying the stationary phase approximation as discussed in \cite{ks1} (see also \cite{ks2}). Assuming $\b_k\ll1$, \eq{ab} can be solved iteratively by using the initial condition \eq{ic}. The {\it first order} solution\footnote{One may see from \eq{ab} that $\a_k\simeq 1+{\cal O}(\b_k^2)$.} is given by
\be
\a_k(t)\simeq 1,\hs{5}\b_k(t)\simeq \int_{t_i}^t\,dt'\fr{\dot{\o}_k(t')}{2\o_k(t')}e^{-2i\th_k(t')}.\label{absol}
\ee
For $\o_k\gg H$, the phase $e^{i\th_k(t')}$ oscillates very rapidly in time $t'$. Furthermore, from \eq{o} and \eq{back} one finds
\bea
\dot{\o}_k(t')&=&\fr{1}{2\o_k(t')}\left[ -2H\fr{k^2}{a^2}+\s\dot{\F}\sin(mt')  +\s m\F\cos(mt')\right]\nn\\
&\simeq&\fr{1}{2\o_k(t')}\left[ -2H\fr{k^2}{a^2}+\fr{\s m\F}{2}(e^{imt'}+e^{-imt'})\right],\label{od}
\eea
where in the second line  $\dot{\F}\sim H\F$  is neglected compared to $m\F$  since $m\gg H$. Using \eq{od} in \eq{absol}, one sees that the first and the third terms still give highly oscillating integrands that are negligible. On the other hand, the phase of the second term is given by $mt'-2\th_k(t')$, which is stationary at time $t_k$  defined by the relation $2\o_k(t_k)=m$. Since we take $m^2\gg \s \F_i$, see \eq{c1}, and we assume $m\gg \mm$, the time  $t_k$ can be fixed as
\be\label{tk}
\fr{k}{a(t_k)}= \fr{m}{2}. 
\ee
For a given $k$, $\b_k(t)$ is vanishingly small for $t<t_k$. For $t>t_k$, the integral in \eq{absol} can be evaluated using the stationary phase approximation. From the second time derivative of $mt'-2\th_k(t')$ evaluated at $t_k$, one sees that the approximation is applicable if $\dot{\o}_k(t_k)/H^2\gg1$.  To satisfy this condition for all times, we require $m^2\gg \s M_p$, which is the second condition stated in \eq{c1}. After these considerations, it is easy to employ  the stationary phase approximation that yields 
\be\label{bs}
\b_k(t)\simeq \left\{ \begin{array} {ll} \fr{2\s\sqrt{\F_i M_p}}{m^{5/4}}(2k)^{-3/4}\exp[imt_k-2i\th_k(t_k)+i\pi/4]\hs{5}\fr{m}{2}< k<\fr{a(t)m}{2}.\\ \\
0 \hs{74}\textrm{otherwise},
\end{array}\right. 
\ee
where we have used $6^{1/4}\sqrt{\pi/2}\simeq 2$. For this whole process to be consistent, one must have $|\b_k|\ll1$ and from \eq{bs} this gives the first condition in \eq{c1}. 

The above solution can be thought to describe the decay of the inflaton with mass $m$ to two $\chi$-particles with momenta $m/2$ as fixed by \eq{tk} \cite{ks1,ks2}.  From the standard interpretation of the Bogoliubov coefficients, $|\b_k|^2/(2\pi)^3$ gives the number density of the created modes with momentum $k$. As a result, the energy density of the created $\chi$-particles can be found as
\be\label{rk}
\rho_\chi=\fr{1}{(2\pi)^3}\int d^3k |\b_k|^2\o_k.   
\ee
The backreaction effects start when this energy density catches up the background inflaton energy density. The corresponding time $t_f$ can be found from the condition
\be\label{tf0}
\rho_\chi(t_f)=\fr12 m^2 \F(t_f)^2.
\ee
To simplify the formulas below, we assume  that for all momenta in the decay range one has 
\be\label{wka}
\o_k\simeq \fr{k}{a}:\hs{7} \fr{m}{2}<k<\fr{a(t_f)m}{2}.
\ee
For this last condition to hold $\F_i$ should not be too large. Indeed, to satisfy \eq{wka} for all times in the interval $(t_i,t_f)$ and for all momenta in the decay range $m/2<k<a(t_f)m/2$ one must take\footnote{To be more precise, the condition \eq{fcond} ensures that $\o_k\simeq\sqrt{k^2/a^2+\mm^2}$. However, since $m\gg \mm$ the presence of $\mm$ in $\o_k$ does not change our estimates  too much.}  
\be\label{fcond}
m^2> 4\s \F_i\left(\fr{t_f}{t_i}\right)^{1/3}.
\ee
Under this assumption, using \eq{bs} and \eq{wka} in \eq{rk} one sees that 
\be
\rho_\chi(t)\simeq \fr{\s^2M_p}{20\pi^2}\F(t)\left(1-\fr{1}{a(t)^{5/2}}\right).
\ee
From \eq{tf0}, the value of the inflaton amplitude $\F_f$ just before the backreaction sets in can be found as in \eq{pf}. As noted above, the corresponding time $t_f$ can be determined from \eq{t}. 

In evaluating the loop corrections numerically, we will use the following set for the parameters of the model
\be
m=10^{-6}M_p,\hs{5}\s=10^{-13}M_p.\label{ns}
\ee
In the chaotic $m^2\f^2$ model, the inflaton mass is fixed as in \eq{ns} by the amplitude of the scalar metric perturbations and we use the same value for convenience. Note that \eq{ns} obeys the second condition in \eq{c1}. From \eq{pf} one also finds 
\be
\F_f\simeq10^{-16}M_p.
\ee
The  reheating temperature is fixed by \eq{rt} as $T_r\simeq 10^{-11}M_p$ and the corresponding Hubble parameter can be found as $H(t_f)\equiv H_f=10^{-22}M_p$. On the other hand, using \eq{ns} in \eq{fcond}  gives $\F_i<10^{-4} M_p$. Together with this condition, \eq{wka} is satisfied for the numerical set \eq{ns}.  As a result, in our estimates below we choose 
\be\label{fi}
\F_i=10^{-4}M_p. 
\ee
Note that \eq{ns} and \eq{fi} obey the inequality indicated in the footnote \ref{foot1}. Using \eq{t}, the corresponding value of the initial Hubble parameter can be found as $H_i\equiv H(t_i)\simeq 10^{-10}M_p$.

\section{Loop corrections to cosmological correlations} \label{s4}

In this section, we calculate the 1-loop corrections to the scalar and the tensor power spectra\footnote{Note that due to the homogeneity and the isotropy of the background, the tadpole $\lf \vf(t,\vec{x})\rg$ becomes a function of time only and thus it can be viewed as a (presumably) small correction to the background evolution.} that arise by the $\chi$-modes circulating in the loops.\footnote{In general, one may also calculate the loops of $\vf$ and $\cc_{ij}$, but since the $\chi$-background vanishes in our period of interest these corrections are identical to the ones obtained in the single scalar field models, which have been extensively studied in the literature showing that no significant contribution can arise after horizon crossing.} As shown in \cite{w1}, the vacuum expectation value of a given operator $O$ can be determined order by order using the in-in formalism as  
\be\label{inp}
\left< O(t)\right>=\sum_{N=0}^{\infty} i^N \int_{t_i}^t dt_N\int_{t_i}^{t_N}dt_{N-1}...\int_{t_i}^{t_2}dt_1\left< [H_I(t_1),[H_I(t_2),...[H_I(t_N),O(t)]...]\right>,
\ee
where $H_I$ is the interaction Hamiltonian in the interaction picture. As we will see below, for the loop corrections of our interest \eq{inp} reduces to the time integrals of the $\chi_k$-mode functions. During preheating, these mode functions exponentially grow that makes the loop corrections meaningfully large \cite{a1,a2}. In perturbative reheating, the mode functions do not enlarge but the process  takes a lot longer giving larger integration ranges for the quantum corrections in \eq{inp}. Thus, one would expect an enhancement for non-oscillating integrands (as we will see, in the present model this expectation is only partially fulfilled since the integrands become oscillatory). 

By expanding the full action \eq{a} around the background solution \eq{back}, one may obtain an action that is ordered in the number of field fluctuations. Since the fluctuations are assumed to be small, the largest corrections presumably arise from the cubic interactions. In our case, the cubic interactions that involve two $\chi$-fields can be determined as   
\be
S^{(3)}_{\chi\chi}=\int a^3 \left[-\fr{\s\f\df}{4HM_p^2}\vf\chi^2-\fr{\df}{4HM_p^2a^2}\vf(\del\chi)^2-\fr{\df}{4HM_p^2}\vf\dot{\chi}^2-\dot{\chi}N^i\del_i\chi-\fr{\s}{2}\vf\chi^2+\fr{1}{a^2}\cc^{ij}\del_i\chi\del_j\chi\right].\label{s33}
\ee
The last term containing the graviton $\cc_{ij}$ modifies the tensor power spectrum. Other interactions are relevant for the scalar power spectrum and we see that all but one of them are suppressed by $M_p^2$. 

\begin{figure}
\centerline{
\includegraphics[width=6cm]{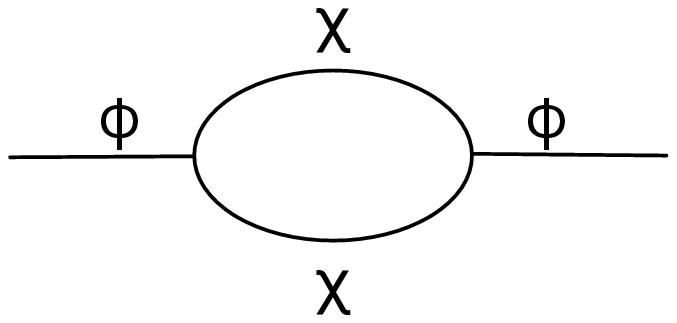}\hs{5}\includegraphics[width=6cm]{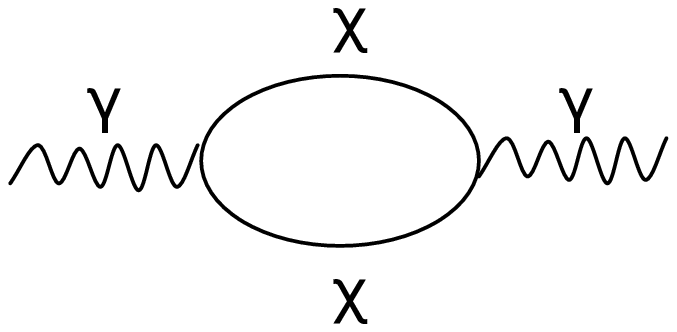}}
\caption{The schematic picture of the 1-loop graphs arising from the interaction Hamiltonians \eq{h1} and \eq{h2} that contribute to the scalar and the tensor power spectra,  $\lf\vf\vf\rg$ and $\lf\cc_{ij}\cc_{kl}\rg$, respectively.}
\label{fig1}
\end{figure}

For the loop corrections to the scalar power spectrum, we first focus on the next to the last term involving the coupling $\s$ since it is not suppressed by the Planck mass (later we show that  $\vf(\del\chi)^2$ term gives a larger contribution). The corresponding interaction Hamiltonian is given by 
\be\label{h1}
H_I=\int d^3 x\, a^3\, \fr{\s}{2}\vf\chi^2.
\ee
Using \eq{h1}  in \eq{inp} for $O=\vf\vf$ with $N=2$ gives the following 1-loop correction to the scalar power spectrum:
\bea
P_k^\vf(t)^{(1)}=&&\fr{\s^2}{(2\pi)^3}\int_{t_i}^t dt_1\int_{t_i}^{t_1}dt_2\,\int d^3q\,a(t_1)^3\,
a(t_2)^3 \label{1} \\
&&\left[\chi_q(t_1)\chi_{k+q}(t_1)\chi_{q}^*(t_2)\chi_{k+q}^*(t_2)\right] \vf_k(t)\vf_k^*(t_2)\left[\vf_k^*(t)\vf_k(t_1)-\vf_k(t)\vf_k^*(t_1)\right]+c.c. \nn
\eea
which can be pictured as in Fig. \ref{fig1}. From the mode functions of the inflaton field given in \eq{sh}, the leading order contribution for superhorizon $k$ at time $t$ can be found as 
\be
P_k^\vf(t)^{(1)}\simeq i\s^2\int_{t_i}^{t} dt_1\int_{t_i}^{t_1}dt_2\,a(t_1)^3a(t_2)^3\,\fr{\df(t_f)^2}{H(t_f)^2}
\fr{\df(t_1)}{H(t_1)}\fr{\df(t_2)}{H(t_2)}\left[f(t)-f(t_1)\right]\CF\, |\zo|^2,   \label{11}
\ee
where 
\be\label{F}
\CF=\fr{1}{(2\pi)^3}\int  d^3q\,\left[\chi_q(t_1)\chi_{k+q}(t_1)\chi_{q}^*(t_2)\chi_{k+q}^*(t_2)-c.c.\right].
\ee
Note that the complex function $\CF$ depends on  $t_1$, $t_2$ and the external superhorizon momentum $k$. Eq. \eq{11} modifies the tree level scalar power spectrum given in \eq{3l}. 

Similarly, the last term in \eq{s33} gives the following interaction Hamiltonian 
\be\label{h2}
H_I=-\fr12 \int d^3 x\, a \,\cc_{ij}\del_i\chi\del_j\chi. 
\ee
Using \eq{h2}  in \eq{inp} for $O=\cc_{ij}\cc_{kl}$ with $N=2$ gives the 1-loop correction that can be pictured as in Fig. \ref{fig1}. From the superhorizon graviton mode functions in \eq{sh}, the corresponding leading order  correction at superhorizon scales at time $t$ can be found as  
\be\label{22}
P_k^\cc(t)^{(1)}\simeq \fr{i}{M_p^2}\int_{t_i}^{t}dt_1\int_{t_i}^{t_1}dt_2\, a(t_1)\,a(t_2)\,\left[g(t_1)-g(t)\right]\CH|\cco|^2,
\ee
where 
\be\label{H}
\CH=\fr{1}{(2\pi)^3}\int d^3q \,(q_\perp^2)^2\,\left[ \chi_{q}(t_2)\chi_{k+q}(t_2)\chi_{q}^*(t_1)\chi_{k+q}^*(t_1)-c.c.\right]
\ee
and $q_{\perp}$ is the  part  of $q$ that is perpendicular to $k$, i.e.  $(q_{\perp})_i=P_{ij}(k)q_j$. This correction modifies the tree level tensor power spectrum given in \eq{3l}. 

Not surprisingly, the functions $\CF$ and $\CH$ given above are divergent and the loop corrections must  be  regularized. The degree of divergence of each function can be found by noting that $\chi_q\sim 1/\sqrt{q}$ as $q\to\infty$, i.e. the Bunch-Davies mode functions approach to the flat space counterparts at large momenta. The dimensional regularization is impossible to utilize here since the exact analytic  form of the mode function $\chi_q$ is not known. Another possible approach is to use the Pauli-Villars regulator fields together with the WKB approximation, as discussed in \cite{wpv}. Here, we generalize the well known adiabatic regularization technique of \cite{ad1,ad2} to the in-in loop integrals, which becomes analogous to the WKB approximation used in \cite{wpv}. A crucial point to remember is that the model at hand involves gravity and it is non-renormalizable. Therefore, one should find a natural way of fixing the finite parts of the loop integrals after infinities are subtracted. In the adiabatic renormalization prescription that we employ below, the finite loop contributions are associated with the particle creation effects and  they are uniquely determined. In other words, the finite parts are switched on by the particle creation effects on the time dependent backgrounds  and they vanish when the time dependence is turned off. 

The adiabatic renormalization prescription can be utilized as follows: One may use the WKB  mode function \eq{wkb} in the divergent expressions like \eq{F} and \eq{H}, and group the terms according to the number of $\a_q$ and $\b_q$ coefficients that they contain. From the equations of motion \eq{ab}, it is possible to see that $\a_q\to1$ and $\b_q\to0$ as $q\to\infty$. Indeed, one has $\b_q\sim e^{-2iq\eta}$, where $\eta$ is the conformal time defined by $d\eta=dt/a$, and with the  $i\e$ prescription necessary to define the Bunch-Davies vacuum state at $\eta=-\infty$ (see e.g. \cite{mal}), $\b_q$ vanishes exponentially at large momenta. As a result, in a loop integral the term containing only  $\a_q$ coefficients  diverges; and all others containing at least one $\b_q$ factor converge. For renormalization, it is then enough to throw out this term that has only $\a_q$ factors. Since $\b_q$ coefficient is associated with the particle creation effects, the remaining finite loop contribution can be thought to be produced by the particles created out of vacuum. In this way, the finite part of the divergent loop correction is fixed uniquely. 

In our case, $\b_q$ is significant only in the finite interval given by \eq{bs} and we have $q_{phys}={\cal O}(m)$. Thus, for $k$ being the cosmological scale of interest, one has $k\ll q$ and to a very good approximation the $k$ dependence in \eq{F} and in \eq{H} can be neglected.\footnote{In \eq{H}, choosing $k$ along the $q_z$ axis one has $q_\perp= q\sin(\th)$. Next, integrating over $\th$ gives a factor close to unity and thus one may take $q_\perp\simeq q$.} Since, we also have $|\a_q|\simeq1$ and $|\b_q|\ll1$, the leading order contributions arise from the $\a_q^3\b_q$ terms after the $\a_q^4$ term is  thrown away for regularization. 

Using the background solution \eq{back} one may see that the time $t_k$ defined in \eq{tk} becomes
\be
t_q\simeq t_i\left(\fr{2q}{m}\right)^{3/2},
\ee
and phase integral defined in \eq{o} is given by 
\be
\th_q(t)\simeq 2q\left[\fr{1}{H(t)a(t)}-\fr{1}{H_i}\right],
\ee
where $H_i=H(t_i)$. Then, the leading order finite parts of $\CF$ and $\CH$ can be seen to involve the following integral
\be\label{int1}
\int_{m/2}^{a(t_1)m/2}\,dq\, (q^n)\,\sin\left[r(q,t_2)\right]-(t_1\leftrightarrow t_2)
\ee
where 
\be\label{rq}
r(q,t)=-\fr{4m}{3H_i}\left(\fr{2q}{m}\right)^{3/2}+\fr{4q}{H(t)a(t)}+\fr{\pi}{4}
\ee
and $n=-3/4$ for $F$ and $n=13/4$ for $H$. From \eq{11} and \eq{22}, one sees that $t_1>t_2$. 

Before estimating the integral \eq{int1}, let us note the overall scale factor dependencies of our corrections. The loop integrals in \eq{F} and \eq{H} are restricted in the range $a(t_i)m/2<q<a(t_f) m/2$ and thus the measure $d^3 q$ would yield a factor $a^3$. Since $\chi_q\propto 1/a^{3/2}$, we have $\CF\propto 1/a^3$ and $\CH\propto a$. Furthermore, we also have $f\propto 1/a^3$ and $g\propto 1/a^3$, which shows that  
overall the number of scale factors are canceled  out in the loop corrections \eq{11} and \eq{22}, i.e. the corrections are invariant under the scaling $a\to \l a$, as they should be. As a result, the suppression of the individual $\chi$-modes during inflation is compensated and one can safely set $a(t_i)=1$ in the formulas. 

The integrand in \eq{int1} is highly oscillatory but fortunately the phase is stationary in one of the integration regions. From the phase given in \eq{rq}, the stationary point can be found as
\be\label{stp}
q_*=\fr{a(t)m}{2}.
\ee
In \eq{int1}, the corresponding stationary point is in the integration range of the first integral and it is located outside of the second one since $t_1>t_2$.  Therefore, the second integral is negligible and applying the stationary phase approximation to the first one yields
\be\label{gf}
\int_{m/2}^{a(t_1)m/2}\,dq\, (q^n)\,\sin\left[r(q,t_2)\right]-(t_1\leftrightarrow t_2)\simeq \fr{\sqrt{\pi}}{2^{n+1/2}}\,\sqrt{H_i}\,m^{n+1/2}\,a(t_2)^{n+1/4}\,\sin[mt_2].
\ee
Using this formula with $n=-3/4$ for $\CF$ and with $n=13/4$ for $\CH$, one obtains
\bea
&&\CF\simeq \fr{i}{\sqrt{2}\pi^{3/2}}\fr{1}{a(t_1)^2 a(t_2)^{5/2}}\fr{\s\sqrt{\Phi_i H_i M_p}}{m^{3/2}}\sin(mt_2),\label{FH1}\\
&&\CH\simeq \fr{i}{16\sqrt{2}\pi^{3/2}}\fr{a(t_2)^{3/2}}{a(t_1)^2 } m^{5/2}\s\sqrt{\Phi_i H_i M_p}\sin(mt_2).\label{FH2}
\eea
These are properly renormalized expressions, which are ready to be used in \eq{11} and \eq{22}, respectively. 

From \eq{FH1}-\eq{FH2} one may  straightforwardly calculate the 1-loop corrections \eq{11} and \eq{22}, which involve elementary integrals. To proceed, we observe from \eq{fgs} that $f$ and $g$ are decreasing functions of time with certain powers and $f(t_1)\gg f(t)$ and $g(t_1)\gg g(t)$, except in a comparatively small integration region where $t_1$ approaches to $t$ (note that we have $t_f\gg t_i$). After neglecting $f(t)$ and $g(t)$ terms in \eq{11} and \eq{22}, one sees that the leading order corrections become proportional to tree level results\footnote{Although they are negligible at superhorizon scales, the loop corrections \eq{11} and \eq{22} have nontrivial momentum dependencies, which are completely different than the tree level results. Moreover, the proportionality factors relating loop corrections to the tree-level results become time dependent functions. Finally, we have already utilized a renormalization prescription that uniquely fixes the loop contributions. As a result, it is not possible to absorb the corrections \eq{11} and \eq{22} by wave-function renormalizations.} and in the following we determine their relative magnitude. Using \eq{FH1}-\eq{FH2} in  \eq{11} and \eq{22}, one encounters elementary integrals of the form $\int dt \sin(mt)\,t^b$, for some power $b$. Because in our integration domain $mt\gg1$, the amplitude $t^b$ slowly changes compared to the rapidly oscillating factor $\sin(mt)$. In that case one may use 
\be\label{oi}
\int  dt  \sin(m t)\,t^b=\fr{t^b}{m}\left\{-\cos(m t)+{\cal O}\left(\fr{1}{mt}\right)\right\}.
\ee
As a result, the loop corrections become oscillating functions of time. This is not surprising, at least for the scalar power spectrum, since the tree-level amplitude is already given by an oscillating function in \eq{3l}. The magnitude of the correction depends on the sign of $b$. For $b<0$, the oscillating integral gets its largest contribution from the first cycles and its dependence on $t_f$ becomes negligible. On the other hand, for $b>0$ the correction becomes larger and larger as time increases, and the dependence on $t_i$ becomes negligible. When $b=0$, the integral is oscillatory with constant amplitude. 

For the scalar power spectrum, using \eq{FH1} in \eq{11} and keeping the leading order term\footnote{In finding the leading order term in \eq{sls} and in other expressions below, we check that the neglected terms $f(t_1)$ and $g(t_1)$ in \eq{11} and \eq{22} indeed yield sub-leading corrections. Moreover, we also check that the sub-leading oscillating factors in $H$, that arise from the background Friedmann equation $H^2=\fr{1}{6M_p^2}(\df^2+m^2\f^2)$ also give contributions that are much smaller.}  gives 
\be\label{sls}
P_k^\vf(t)^{(1)}\simeq\fr{1}{\sqrt{2\pi^{3}}}\s^3\sqrt{\fr{\F_iM_p}{H_im^9}}\left[ \fr{\cos(mt)}{t}\right]\,P_k^{\vf(0)}(t).
\ee
This  corresponds to $b=-1$  in \eq{oi} and we see that the contribution decreases as $t\to t_f$. Consequently, the strength of the correction relative to the tree-level result is largest when $t\sim t_i$ and it is fixed by the following dimensionless factor
\be\label{cs1}
\s^3\sqrt{\fr{\F_iH_iM_p}{m^9}}\simeq \fr{\s^3\F_i}{m^4}, 
\ee
where we have used \eq{t} to express $H_i$ in terms of other parameters.  For our numerical set of parameters given in \eq{ns} and \eq{fi}, this factor becomes $10^{-19}$. 

On the other hand, using \eq{FH2} in \eq{22}  and again keeping the leading order term yields 
\be\label{gls}
P_k^\cc(t)^{(1)}\simeq \fr{1}{24\sqrt{2\pi^{3}}}\fr{\s}{M_p^2}\sqrt{\fr{m\F_i M_p}{H_i}}\,\sin(mt)\,P_k^{\cc(0)}.
\ee
This case corresponds to $b=0$ in \eq{oi}, where the sine function is replaced by the cosine. As noted above, for $b=0$ the resulting correction has constant amplitude independent of time $t$. From \eq{gls}, the relative strength of the loop correction as compared to the tree-level result is determined by the following dimensionless combination of the parameters:
\be\label{cs2} 
\fr{\s}{M_p^2}\sqrt{\fr{m\F_i M_p}{H_i}}\simeq \fr{\s}{M_p},
\ee
where we again use \eq{t} for $H_i$. For the numerical set \eq{ns}  this factor equals $10^{-13}$. 

An unexpected feature of the above results is that the relative strength of the tensor power spectrum correction becomes larger than the scalar one, i.e. the factor \eq{cs2} is larger than \eq{cs1}, although \eq{cs2} is suppressed by $M_p$. Tracing back how these are obtained, we see that the main difference between the two arises due to the distinct mass $m$ dependencies in \eq{FH1} and \eq{FH2}. Moreover, while the amplitude of the correction \eq{sls} decreases with time, the amplitude in \eq{gls} is constant. It turns out that the source for these two differences is the same, i.e.  the extra momentum factor appearing in \eq{H} as compared to \eq{F},  which gives both the additional factors of $m$ and the extra factors of time  when the momentum integral is evaluated at the stationary point \eq{stp}, see \eq{gf}. We observe that the momentum pre-factor in \eq{H} appears due to the partial derivatives acting on the graviton field $\cc_{ij}$  in \eq{h2},  and this suggests that  the cubic coupling $\vf(\del\chi)^2$ in \eq{s33} might give a larger correction to the scalar power spectrum  by the same mechanism, although it is suppressed by $M_p$.

To see whether this is the case or not, one can use the corresponding interaction Hamiltonian 
\be\label{h3}
H_I=\int d^3 x\, \fr{a\df}{4HM_p^2}\vf(\del\chi)^2
\ee
in \eq{inp}. Then, a straightforward calculation gives the following 1-loop correction
\be
P_k^\vf(t)^{(1)}\simeq \fr{i}{4M_p^4} \int_{t_i}^{t} dt_1\int_{t_i}^{t_1}dt_2\,a(t_1)a(t_2)\,\fr{\df(t_f)^2}{H(t_f)^2}
\fr{\df(t_1)^2}{H(t_1)^2}\fr{\df(t_2)^2}{H(t_2)^2}\left[f(t)-f(t_1)\right]\tilde{\CF}\, |\zo|^2,   \label{111}
\ee
where 
\be\label{TF}
\tilde{\CF}=\fr{1}{(2\pi)^3}\int  d^3q\,q^2(q+k)^2\,\left[\chi_q(t_1)\chi_{k+q}(t_1)\chi_{q}^*(t_2)\chi_{k+q}^*(t_2)-c.c.\right].
\ee
Note that this correction is obtained by using  \eq{h3}  in \eq{inp} for $O=\vf\vf$ with $N=2$ and it can still be pictured as in Fig \ref{fig1}. When $k$ is the superhorizon scale of interest, one sees that 
\be
\tilde{\CF}\simeq \CH.
\ee
From \eq{FH2} and \eq{111}, it is now straightforward to obtain the following leading order correction
\be\label{glss}
P_k^\cc(t)^{(1)}\simeq \fr{3}{5(16)^2\sqrt{2\pi^{3}}}\fr{\s\, m^{3/2}}{M_p^{4}}\sqrt{ H_i\F_i^5M_p}\,\left[\cos(mt)^5\,t_i^{5/3}\,t^{1/3}\right]\,P_k^{\cc(0)}.
\ee
This time, the power of $t_1$ in the integral in \eq{111} becomes positive and the relative strength of the correction increases with time. As a result, the correction is maximized for $t\sim t_f$ whose relative magnitude is determined by the dimensionless factor
\be\label{cs3}
\fr{\s\,\sqrt{m^{3}\F_i^{5}M_p}}{M_p^{4}H_i^{7/6}H_f^{1/3}},
\ee
which has $H_f$, the smallest mass scale in the problem, appearing in the denominator. For the numerical set of parameters given in \eq{ns} and \eq{fi}, this factor becomes $10^{-13}$, which is 6 orders of magnitude larger than \eq{cs1}. Note that while \eq{cs3} is suppressed by $M_p^4$, the factor characterizing the tensor correction \eq{cs2} is suppressed by $M_p^2$. The first of these directly comes from the interaction Hamiltonian \eq{h3}, which is used twice in the in-in formula \eq{inp}, and the second factor arises from the normalization of the graviton mode function given in \eq{w}. 

It is possible to examine \eq{cs1} and \eq{cs3} by dimensional analysis. Ignoring the differences between the initial and the final values of the time dependent quantities, \eq{cs1} and \eq{cs3} become $\s^3\F/m^4$ and $\s m^2\F^3/(H^2M_p^4)$, where we have used \eq{t} to simplify \eq{cs3}. Eq. \eq{cs1} arises from the interaction \eq{h1}, which is used twice, therefore the strength of the correction can be identified as $\s^2$. Similarly, \eq{cs3} arises from \eq{h3}  whose respective strength can be determined as $(p^2m\F/(HM_p^2))^2$, where $m\F$ comes from $d\f/dt$ and $p^2$ denotes the contribution of the derivatives. By comparing these expressions, one sees that $p\sim m$, i.e. a spatial derivative acting on $\chi$ has the strength $m$, which is not surprising since the $\chi$ modes are created by the decay of the inflaton with mass $m$. Although this naive estimate cannot fully account for the differences between \eq{cs1} and \eq{cs3} (because they involve nontrivial momentum and time integrals), it indicates that the interactions containing derivatives of $\chi$ have a suppression factor $m/M_p$ for each derivative. Yet, one should also keep in mind that there are other factors affecting the strength of an interaction like the term $1/H$ appearing in \eq{h3} (recall that $H$ is one of the smallest mass scales in the problem). In any case, the corrections due to the cubic interactions turn out to be unobservably small. By dimensional reasons, the higher order interactions must be suppressed by smaller and smaller combinations of parameters (since otherwise the expansion of the action around the classical background solution would fail from the beginning), the entropy mode loop corrections to the cosmological correlation functions can safely be ignored in this model. 

On the other hand, one may also compare these corrections to the ones arising from $\chi$-loops during inflation. If one ignores the possible existence of infrared logarithms, the relative magnitude of the corrections during inflation are suppressed by $H^2/M_p^2$, where $H$ is the corresponding Hubble parameter, see e.g. \cite{ekw}. For our numerical set of parameters $H_i=10^{-10}M_p$ and thus the dimensionless number characterizing such corrections is $10^{-20}$, which is much smaller than the analogous contributions \eq{cs2} and \eq{cs3} arising in perturbative reheating. 

\section{Conclusions} \label{s5} 

In this paper, we calculate the entropy mode loop corrections to the scalar and the tensor power spectra during perturbative reheating. These corrections arise during the first stage of reheating, which starts just after inflation and ends when the backreaction effects become important. In that period, the homogenous and isotropic background corresponds to the coherently oscillating inflaton field, and the quantum excitations of the field perturbations can still be treated as small fluctuations, where the in-in perturbation theory can be used to treat non-linear effects arising from interactions. 

The model we studied has two scalars: the inflaton $\f$ and the reheating scalar $\chi$. While the inflaton fluctuations correspond to the adiabatic modes, the $\chi$-fluctuations become entropy perturbations. During reheating, $\chi$-modes are exited by the oscillating inflaton background and we determine their contribution to the cosmological correlations via the loop effects using in-in perturbation theory. For that, we focus on the cubic interaction terms in the Lagrangian involving two $\chi$-scalars. These interactions give rise to the 1-loop corrections to the scalar and the tensor power spectra, which can be pictured as in Fig. \ref{fig1}. It turns out that the strengths of these entropy mode loop corrections  depend on the various  parameters of the model in a nontrivial way, see  \eq{cs1}, \eq{cs2} and \eq{cs3}. The main input parameters are the inflaton mass $m$, the cubic coupling constant $\s$ that fixes the decay rate of the inflaton to the two $\chi$-particles, and the initial inflaton amplitude $\F_i$ in the beginning of reheating. The other parameters, i.e. $H_i$, $H_f$ and $\F_f$, can be determined in terms of the main input variables. The loop corrections turn out be small, especially when they are compared to the preheating case \cite{a1,a2}, but they are still much larger than the analogous corrections that arise during inflation \cite{ekw}.   

To regularize/renormalize the divergent loop integrals, we adapt the well known adiabatic regularization technique of \cite{ad1,ad2} to the in-in formalism. This adiabatic renormalization scheme naturally regularizes the loops, unambiguously fixes the finite parts of the divergent integrals and offers a viable alternative to the dimensional or Pauli-Villars regularizations in cosmology.  It would be interesting to develop/interpret this procedure in terms of the standard renormalization procedure by identifying the counterterms added to the  action that absorb infinities. It would also be interesting to compare it with other renormalization schemes for the well studied cases like a self-interacting scalar field in de Sitter space.  On the other hand, in the present model the time integrals arising in the main formula \eq{inp}  become oscillatory that diminishes the magnitude of the correction. This is contrary to the preheating case studied in \cite{a1,a2}. It is of interest  to find out models where the time integrals in \eq{inp} are non-oscillatory, which would presumably yield larger loop corrections.

\end{document}